# Photoluminescence of silicon-vacancy defects in nanodiamonds of different chondrites

A. A. SHIRYAEV[1, 2, #], A. V. FISENKO[3], L. F. SEMJONOVA[3], A. A. KHOMICH[4],

and I. I. VLASOV[4,5]

[1] Institute of Physical Chemistry and Electrochemistry RAS, Leninsky pr. 31, korp. 4, 119071, Moscow, Russia

[2] Institute of Ore Deposits, Petrography, Geochemistry and Mineralogy RAS, Staromonetny per. 35, 119071, Moscow, Russia

[3] Vernadsky Institute of Geochemistry and Analytical Chemistry RAS, Kosygin Street 19, Moscow, Russia

[4] General Physics Institute RAS, Vavilov Street 38, 119991 Moscow, Russia

[5] National Research Nuclear University MEPhI, Kashirskoe Avenue 31, Moscow 115409, Russia

[#] Corresponding author email: shiryaev@phyche.ac.ru AND a_shiryaev@mail.ru

**Abstract**

Photoluminescence spectra show that silicon impurity is present in lattice of some nanodiamond grains (ND) of various chondrites as a silicon-vacancy (SiV) defect. The relative intensity of the SiV band in the diamond-rich separates depends on chemical composition of meteorites and on size of ND grains. The strongest signal is found for the size separates enriched in small grains; thus confirming our earlier conclusion that the SiV defects preferentially reside in the smallest (≤ 2 nm) grains. The difference in relative intensities of the SiV luminescence in the diamond-rich separates of individual meteorites are due to variable conditions of thermal metamorphism of their parent bodies and/or uneven sampling of nanodiamonds populations. Annealing of separates in air eliminates surface $sp^2$-carbon, consequently, the SiV luminescence is enhanced. Strong and well-defined luminescence and absorption of the SiV defect is a promising feature to locate cold (< 250 °C) nanodiamonds in space.

**INTRODUCTION**

Nanodiamonds present in meteorites (meteoritic nanodiamonds – MND) remain an enigmatic substance. Their matrix normalized abundance may reach 2000 ppm (Huss and Lewis, 1995). Since their discovery (Lewis et al., 1987) the formation process(es) and astrophysical source(s) of MND remain highly debatable despite significant efforts (see extensive review by Daulton (2005)). Whereas isotopic composition of noble gases and, in particular, of Xe indicates that some of nanodiamonds might be related to supernovae explosions (e.g., Lewis et al., 1987, Jorgenson 1988; Ott et al. 2012), isotopic compositions of bulk carbon and of the principal chemical impurity – nitrogen – in the diamond-rich separates are less conclusive and may support hypothesis of that at least a fraction of MND was formed in the Solar system (Dai et al., 2002). Processes of MND formation are also debatable, but combined analysis of information about structure and chemical impurities (Shiryaev et al., 2011) suggests that the growth process of (at least) N-containing grains should be very fast. The CVD-like (Chemical Vapour Deposition) process (Daulton et al., 1996), possibly triggered by a shock wave(s) (Shiryaev et al., 2011), looks plausible, but other processes cannot be excluded (e.g., Blake et al., 1987; Byakov et al., 1990; Duley and Grishko, 2001; Kouchi et al., 2005; Stroud et al., 2011).

Analysis of isotopic composition of different elements is indispensable for identification of nanodiamonds source(s). Eventual detection of nanodiamonds in astrophysical spectra would provide additional information. Detailed knowledge of spectroscopic properties of meteoritic nanodiamonds is thus of considerable importance. Attempts to observe nanodiamond features in astrophysical spectra are rather numerous, but very few of them can be (relatively) unambiguously assigned to diamonds. The most reliable observations are those by Guillois et al. (1999) and van Kerkhoven et al. (2002) who reported observation of infra-red features at 3.43 and 3.53 microns in spectra of several Herbig Ae/Be stars. Perfect match of these bands to peculiar configuration of C-H bonds on surfaces of "large" nanodiamonds (about 50 nm) makes the assignment of the observed bands to heated nanodiamonds plausible. Subsequent spatially resolved studies (Habart et al., 2004; Goto et al., 2009) showed that the diamond-related emission originates from the inner region (<15 AU) of the circumstellar dust disk, whereas PAH emission extends towards the outer region. The observed spatial heterogeneity may partly reflect temperature and UV flux distribution in the dust disk. However, the "diamond" bands are observed in less than 4% of the studied Herbing stars (Acke and van den Ancker, 2006). It is interesting to note a discrepancy between the sizes of tentatively observed nanodiamonds derived from radiative energy budget (1-10 nm, van Kerkhoven et al. (2002)) and the fact that the required well-resolved IR bands are observed only when the nanodiamond grains become sufficiently large to possess crystal faces resembling those of macroscopic diamond. Investigation of diamond grains smaller than 30 nm shows that the corresponding IR features are blurred and shifted (Sheu et al., 2002; Maturilli et al., 2014). The discrepancy might be resolved if the hydrogen surface coverage is incomplete in contrast to the assumption in van Kerkhoven et al. (2002). Chang et al. (2006) ascribed so-called Extended Red Emission – a broad luminescence-related emission from some astrophysical objects (see Witt, 2013 for review) - to the photoluminescence of nitrogen-vacancy (NV) complexes in nanodiamond particles with sizes approx. 100 nm or larger. Note that the NV defects are extremely difficult to detect in diamond grains smaller than 30-50 nm (e.g., Bradac et al., 2009; Vlasov et al., 2010) and they are not observed in the nanodiamonds extracted from meteorites (Shiryaev et al., 2011).

If correct, these assignments imply existence of large (tens-hundreds of nanometers) nanodiamonds in space. In the same time, the meteoritic nanodiamonds are characterized by considerably smaller sizes (median size ~2.6 nm) as shown independently by TEM (Fraundorf et al., 1989; Lewis et al., 1989; Daulton et al., 1996) and MALDI (Lyon et al. 2005, Maul et al., 2005). Recently published work with atom-probe tomography also showed presence of nanodiamonds in the small size range 2-3 nm (Heck et al. 2014). Spectroscopic properties of nanodiamonds are strongly size-dependent and up to now no reliable astrophysical observations of features resembling spectra of nanodiamonds similar to those from meteorites are known.

Recently we have reported observation of an important point defect – the silicon-vacancy complex (SiV) – in nanodiamond-rich separates from Efremovka (CV3) and Orgueil (CI) chondrites (Shiryaev et al., 2011). Subsequent studies demonstrated that the SiV luminescence appears to be confined to the smallest diamond grains with sizes below 2 nm (Vlasov et al., 2014). We report here on examination of the SiV defects in diamond-rich separates extracted from meteorites of various chemical classes and groups. Implications of these observations for astrospectroscopy are discussed.

## SAMPLES AND METHODS

The diamond-rich separates were extracted from Orgueil (CI), Boriskino (CM2), Efremovka (CV3), Kainsaz (CO3), and Krymka (LL3) chondrites. The extraction was performed according to the standard protocol involving dissolution of the meteorite piece by HCl, HCl+HF, KOH, $H_2O_2$, $K_2Cr_2O_7$, $HClO_4$ at temperatures up to 220 °C (Tang et al., 1988). The separates possess a translucent pale yellow color, typical for meteoritic colloidal diamond. Colloidal ammonia solutions of the Orgueil, Boriskino and Krymka separates were further separated into grain size fractions by centrifugation at various accelerations and duration. Efficiency of the employed separation method is confirmed by the differences in isotopic composition of C, N and noble gases between the different size fractions (e.g., Verchovsky et al., 1998, 2006; Fisenko et al., 2004; Fisenko and Semjonova, 2006).

Photoluminescence (PL) spectra were measured at room temperature for the following samples of nanodiamond-rich separates: bulk separate (OD7) and fine-grain (OD13) fraction (supernatant after 13500 g (50 h) centrifugation of the bulk separate) of Orgueil; BD5 and BD9 - fine- and coarse-grain fractions (supernatant and sediment, respectively) after centrifugation ($10^5$ g, 4h) of Boriskino separate; bulk separate of Kainsaz; DE1 and DE2 – bulk separates of Efremovka, but the sample DE1 was additionally subjected to heavy liquid sedimentation ($\rho=2$ g/cm$^3$), which depleted the specimen in the grains with density below 2 g/cm$^3$; bulk separate (DKr) and coarse grain (DKr4) fraction (sediments after $10^5$ g (4h) centrifugation of the bulk separate) of Krymka.

The PL measurements were performed using a LABRAM HR spectrometer with an $Ar^+$ laser (488 nm). To avoid eventual heating of the samples by probing laser beam a minute amount of the sample was mechanically pressed into pure oxygen-free copper foil. This method allows production of very thin nanodiamond sample. Since its "fluffiness" is greatly reduced, good thermal contact with thermal sink is achieved. The laser power was limited to 1 mW and even prolonged laser irradiation had virtually no influence on recorded spectra, justifying the applied procedure. The accuracy of the determination of the positions of the peaks in the PL spectra was ±0.1 nm. The spectra were corrected for wavelength-dependent sensitivity of CCD detector. For the bulk and coarse-size (DKr and DKr4) fractions of the Krymka separates the measurements were also performed in temperature range from 25 to 450 °C in water-cooled TS 1500 Linkam stage in air. The duration of our heating experiment

(30-45 min for the total cycle; the exposure to high T's was much shorter) was clearly insufficient for noticeable modification of defects structure of the nanodiamonds.

**RESULTS AND DISCUSSION**

Room temperature photoluminescence spectra of the separates are shown in Fig. 1. Note that due to the confinement effect the PL spectra of nanoparticles remain fairly broad even at low temperatures. The spectra are normalised to the maximum of the broad band and are displaced vertically for clarity. The maximum of the broad structureless band lies between 590 and 610 nm. This broad band represents overlapping signals from non-diamond $sp^2$-carbon in the separates and from defects on the surfaces of diamond grains (Iakoubovskii et al., 2000). Variations in its maximum position are due to differences in surface structure of the nanodiamond grains.

A band around 737 nm is observed in all the samples, albeit with widely different relative intensities (see below). This feature is due to the so-called SiV (silicon-vacancy) defect – a silicon ion in a divacancy (can be represented as a rarefied diamond lattice) (for general review see Zaitsev (2001); Shiryaev et al. (2011), Vlasov et al. (2014) provide information about SiV in nanodiamonds). At temperatures exceeding 250°C the SiV band is not observable in the DKr and DKr4 samples with any confidence, which is explained by strong temperature dependence of this PL feature (e.g., Zaitsev, 2001).

After the heating of the DKr and DKr4 samples in air up to 450 °C and subsequent cooling to room temperature the relative intensity of the SiV luminescence increases considerably in comparison with initial spectra. Such behavior is explained by partial elimination of superficial $sp^2$-carbon on surfaces of nanodiamond grains by oxidation during annealing. This hypothesis is supported by Raman spectroscopy – whereas it is notoriously difficult to obtain a decent diamond Raman peak of meteoritic NDs using excitation in visible range due to strong $sp^2$-carbon scattering, after the annealing the diamond Raman peak at 1332 cm$^{-1}$ is easily observed even upon excitation at 473 nm (Fig. 2). As expected for nanodiamonds (e.g., Vul', 2006) a broad shoulder towards smaller wavenumbers is clearly pronounced.

The relative area of the SiV band (calculated as SiV peak area divided by area of the total PL between 490-800 nm) in as-received samples depends on chemical class and group of meteorite and on the nanodiamond grain size (Fig. 3). As shown by us earlier (Vlasov et al., 2014) the SiV luminescence is observed in the grains with sizes ≤2 nm, which is smaller, than the median ND size of 2.6-2.8 nm determined by TEM (Daulton et al., 1996) and MALDI (Lyon, 2005) for nanodiamonds from Murchison (CM2) and Allende (CV3) chondrites. Presumably, procedures applied to Krymka and Efremovka depleted the "bulk" separates in smaller grains, therefore the SiV intensity for the fractions DKr4 and DE1 decreases (Fig. 4). In the same time, for the Orgueil (CI) and Boriskino (CM2) samples the size effect is small and probably not significant (Fig. 3).

The luminescent SiV defect is characterised by certain atomic configuration. For example, a substitutional Si ion in diamond lattice will possess no PL. Therefore at present we cannot estimate the total silicon content in nanodiamond grains from different meteorites. It is shown, that the SiV defect may be formed both by direct incorporation into a growing diamond and by reaction between substitutional Si with vacancies generated, e.g., by irradiation or Si implantation (Iakoubovskii et al, 2001; D'Haenens-Johansson et al., 2011). The presence of the SiV defects in the smallest diamond grains might seem to be counterintuitive, since effect of NDs self-purification from vacancies and impurities is widely discussed (e.g., Barnard and Sternberg, 2007). However, the first principles Density Functional Theory calculations verified by experiments indicate absence of the self-purification in case of SiV defect (Vlasov et al., 2014). It is shown that the SiV formation energies are virtually identical for nanodiamond grains between 1.1 and 1.8 nm in size when the defects are placed in the centre of the particle. When this defect is placed near the surface its formation energy is only marginally (~0.2 eV) lower than the corresponding value in the centre of the grain. Therefore, the driving force for out-diffusion of the Si ion is small resulting in stable SiVs even in very small nanodiamonds.

The reason why the separates enriched by smallest nanodiamonds possess higher relative intensity of the SiV defects is likely due to existence of several populations of nanodiamond grains, differing in origin and thermal stability (e.g., Huss and Lewis, 1994a). These populations may reflect formation is different astrophysical sources (e.g., supernovae and Solar system). In addition, thermal stability of the populations may play an important role. It should be mentioned that so-called stochastic heating of nanograins by individual UV photons may drastically increase temperature of the smallest (<2 nm) grains (see Van Kerckhoven et al., 2002 for ND-related calculations) and their formation and evolution may differ considerably from that of larger grains (Kochanek, 2014).

The sample-dependent variations of the SiV relative intensities presented at Fig. 3 could be explained by (1) differences in amorphous carbon content in the separates, (2) variations in the SiV defects concentration in nanodiamond grains of different meteorites, or by combination of these factors. These possibilities are discussed below.

1) *Differences in amorphous carbon content in the separates*

Transmission electron microscopy (e.g., Bernatowicz et al., 1990, Garvie 2006, Stroud et al., 2011) and NEXAFS and Raman measurements (Shiryaev et al., 2011) reveal that the diamond-rich separates principally consists of nanodiamond grains covered (partly) by $sp^2$-carbon layer and of subordinate amount of other carbonaceous phases (non-graphitising carbons such as glassy carbon, amorphous C, etc). Some of these compounds may result from severe chemical treatment required for MNDs extraction, but at least a part of it is inherited from the MNDs formation process as suggested by electron microscopy (e.g., Garvie and Buseck, 2006) and by processes of entrapment of P3 noble gases in MND (Fisenko et al., 2014 and refs. therein). Potentially these $sp^2$-phases could be responsible for the broad-band PL dominating spectra of the separates. If the PL is indeed caused mostly by the secondary (independent from the nanodiamonds) phases, the trend shown on Fig. 3 would reflect variations in content of non-diamond carbon in the separates due to differences in thermal metamorphism of meteorites parent bodies.

However, studies of behaviour of the broad PL band from various nanodiamonds during oxidation, thermal treatment etc. showed that this band is largely due to $sp^2$-carbon layer on surfaces of nanodiamond grains (e.g, Smith et al., 2010). The present work shows also that annealing of the Krymka separates in air decreases the PL intensity, but subsequent storage in air gradually recovers it, thus supporting its genetic link to $sp^2$-carbon directly connected with nanodiamond grains. Moreover, only one mode of carbon release is observed during step-oxidation of the separates of different meteorites (e.g., Russell et al., 1996), indicating that the fraction of the $sp^2$-carbon in the studied separates is either not abundant or its resistance to thermal oxidation is similar to nanodiamond. The plausibility of the latter assumption remains an open question. Therefore, we prefer another explanation of the observed increase of the SiV relative intensity.

2) *Variations in concentration of the SiV defects in nanodiamond grains of different meteorites.* These variations could be explained by existence of several nanodiamonds populations with strongly variable concentration of the SiV defects. The SiV concentration as well as ratio of the populations in a meteorite depends on conditions of thermal metamorphism of meteorites parent bodies. Recall that the SiV PL in Krymka samples is enhanced by oxidative annealing which partly removes and reconstructs the sp$^2$-carbon superficial layer.

The variations between thermal history of parent meteorites bodies (both temperature and time) may explain the observed spread. Indeed, the lowest intensity of the SiV defects (Fig. 3) are observed for nanodiamonds of Orgueil (CI) and Boriskino (CM2) meteorites which underwent low temperature (~100 °C) metamorphism only (Huss and Lewis, 1994b). The temperature of ~100 °C is too low to induce atomic diffusion in diamond. The SiV is stronger in MND from Efremovka (CV3) and Krymka (LL3.1) meteorites, with temperature of metamorphism of at least 300-400 °C (Huss et al., 2006). In addition, Krymka underwent at least two strong shocks which led to residual meteorite temperature up to 500 °C (Semenenko et al., 1987).

Remarkably, the tendency of changes of the SiV relative intensity shown on Fig. 3 for carbonaceous meteorites correlates with concentration of the P3 component of noble gases in MND, which might be related to disordered carbonaceous phase on surfaces of nanodiamond grains (Fisenko et al., 2014 and refs. therein). In the studied meteorite set the concentration of the P3 noble gases in the NDs of Orgueil and Boriskino is the highest, whereas for the Efremovka NDs it is the lowest (Huss and Lewis, 1994a; Verchovsky et al., 1998; Fisenko et al., 2004). Note also that the maximum of the broad PL band of the Orgueil and Boriskino is somewhat shifted to smaller wavelengths in comparison with other samples. Most likely this shift indicate some differences in the surface sp$^2$-carbon.

Therefore, the enhancement of the SiV luminescence in the diamond-rich separates is largely due to thermal metamorphism (annealing) of meteorite parent bodies, which has promoted defects diffusion, and modification of the surface layer of the nanodiamonds.

**Implications for astrophysical search of nanodiamonds**

The silicon-vacancy defect in meteoritic nanodiamonds is potentially important for understanding origin of nanodiamonds because of several reasons. As shown by Vlasov et al. (2014) up to three emitting centers per particle containing ~400 atoms were observed, though on limited number of grains. If these numbers are statistically valid, this implies that the structural silicon impurity in meteoritic nanodiamonds may represent a new isotopic system for identification of their cosmochemical sources. However, our preliminary Accelerator Mass Spectrometry (Martschini et al., unpublished) and SIMS measurements revealed a problem of contamination of nanodiamonds by silicon: dangling bonds on nanodiamond surfaces may serve as very efficient sorption sites for various elements (e.g., Dolenko et al., 2014). Silicon is fairly ubiquitous element, and appears to be efficiently sorbed by nanodiamonds especially during acid dissolution of silicate matrix of a parent meteorite. Atom Probe Tomography (see Heck et al., 2014) might be the only technique that could discriminate contributions of surface contamination and of nanodiamonds' bulk in Si isotopic analyses.

The luminescence of the SiV defect may also assist in search of astrophysical sources of nanodiamonds with grain sizes comparable with those observed in meteorites. The prominent broad band luminescence of nanodiamonds (see e.g., Fig. 1) is not very useful for such purpose due to several reasons. First of all, the band shape is similar to emission curve of stars belonging to G0-K9 spectral classes and thus unique identification is barely possible. The Extended Red Emission (ERE) is almost never peaked at such wavelengths, being shifted to larger wavenumbers. Only few ERE sources such as galactic cirruses are close in spectral position, but their emission is much narrower (Darbon et al., 1999 and refs. therein). In addition, the broad PL band is related to non-diamond superficial carbon and thus a PL spectrum of a bare nanodiamond in space may differ from that observed in the laboratory. If NDs in stellar environment are less prone to $sp^2$-carbon surface coverage, the relative intensity of the SiV luminescence will be stronger than in the laboratory.

As mentioned in Introduction, Chang et al., (2006) showed a nearly-perfect match of the ERE band from some of the sources with that of the NV defect in large nanodiamonds. However, this resemblance is achieved only in case of very selective narrow band excitation of the NV luminescence. A real stellar source emits quasi-continuum spectrum and in this case several charge states of the nitrogen-vacancy defect will be observed. The overlap of these contributions will make the band too broad to account for the astronomical observations (see Fig. 2 of Chang et al., 2006). In the same time, the SiV defect is characterized by very strong ZPL and the corresponding line at 738 nm remains rather narrow even in nanodiamonds. Strong temperature dependence of the SiV luminescence (e.g., Fig. 2a) permits its observation only from cold particles. However, this defect is also visible in absorption, thus complementing IR observations of C-H bands pronounced in IR emission spectra of hot particles (>400-500 °C).

The search for the SiV luminescence from astrophysical sources is not trivial due to expected weakness of the feature and necessary corrections for sky brightness. One may note a perfect correspondence of the SiV band position with some of Diffuse Interstellar bands (DIBs, Galazutdinov et al., 2000). Unfortunately the DIBs are far too narrow to be positively identified as the SiV luminescence. Bands at similar wavelengths are not uncommon in spectra of supernovae (e.g. Filippenko, 1997), they are mostly due to Doppler-shifted ions emission (such as O[I], Ca[I]) and reliable identification of the SiV requires careful analysis of extended spectral region. A dedicated search for the SiV luminescence and absorption from astrophysical sources is currently underway.

**CONCLUSIONS**

A band of the silicon-vacancy (SiV) defect in diamond lattice is observed in photoluminescence spectra of different grain-size fractions of nanodiamonds (ND) extracted from chondrites of various groups. At present our statistics includes the following classes and groups: CI, CM2, CO3, CV3, and LL3. The concentration of silicon in NDs lattice may reach hundreds of atomic ppm, which makes this element important for identification of astrophysical sources and formation processes of meteoritic nanodiamonds.

The variations in relative intensities of the SiV luminescence between the nanodiamond-rich separates extracted from the meteorites of different classes are due to variations of temperature of thermal metamorphism of their parent bodies and/or uneven sampling of nanodiamonds populations. The silicon impurity is preferentially incorporated into the smallest grains (less than 2 nm). The thermal history of parent meteorite bodies is important for enhancement of the luminescence of trapped Si atoms promoting formation of specific lattice defect and by modification of nanodiamonds surfaces. The strong and well-defined luminescence and absorption of the SiV defect is a promising feature to locate cold nanodiamonds in space.

**Acknowledgments.** This work was partly supported by RFBR grants 12-05-00208 and 15-05-03351. We are grateful to Dr. K. Iakoubovskii and Prof. A. Witt for informative discussions. Detailed reviews of Prof. T. Daulton, P. Heck and I. Franchi were extremely useful.

FIGURES

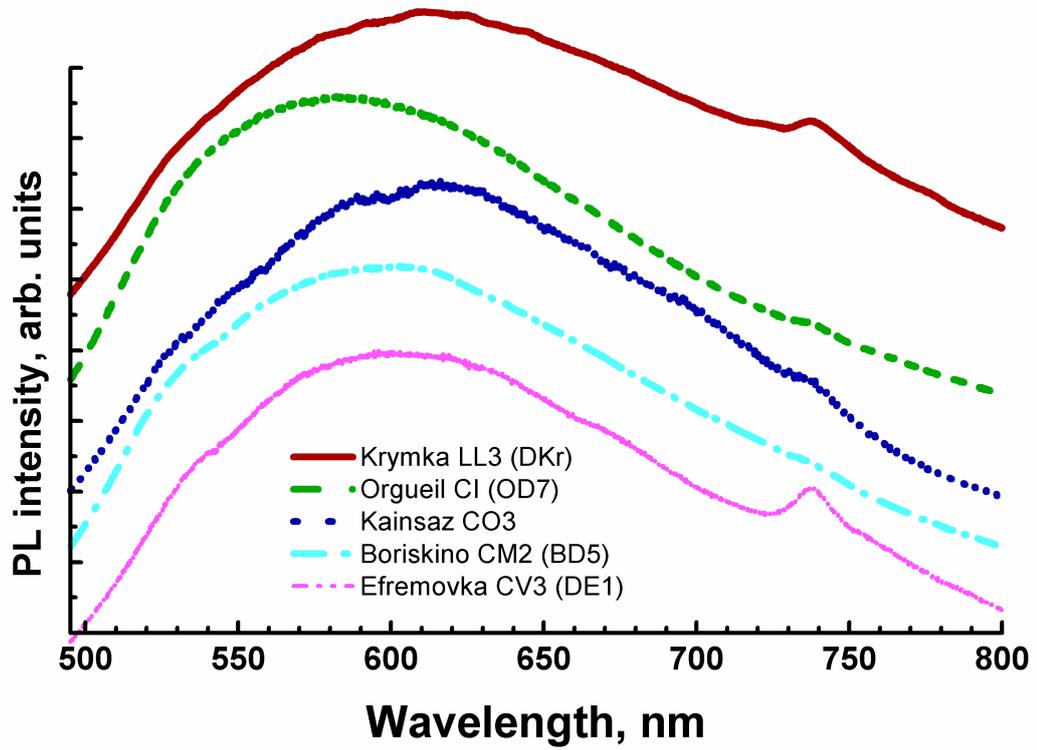

Figure 1. Photoluminescence spectra of studied nanodiamond-rich separates (only some size fractions are shown). The curves are normalized to maximum and displaced vertically for clarity. Note that all shown spectra are related to measurements of the separates in as-received state (without annealing).

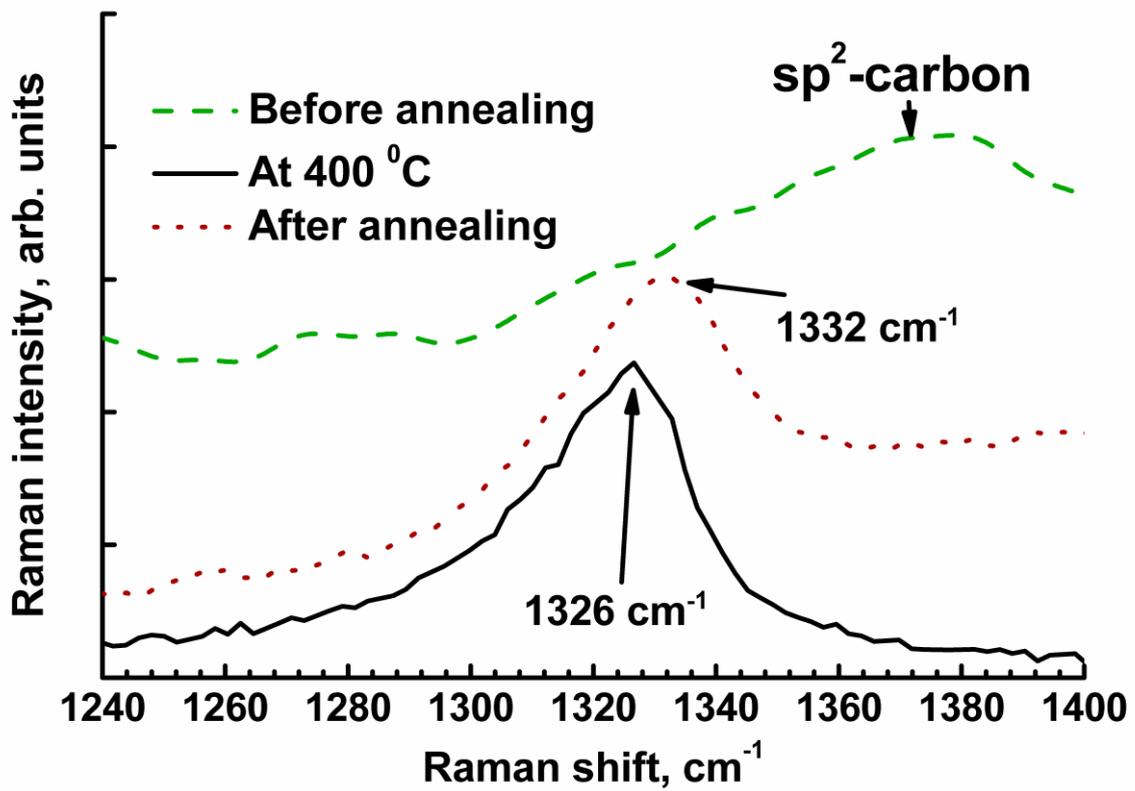

Figure 2. Temperature-induced changes in Raman spectra of the Krymka nanodiamond recorded in as-received state, *in situ* at 400 °C and after the annealing. A linear background partly compensating contribution of temperature-dependent broad-band photoluminescence was subtracted.

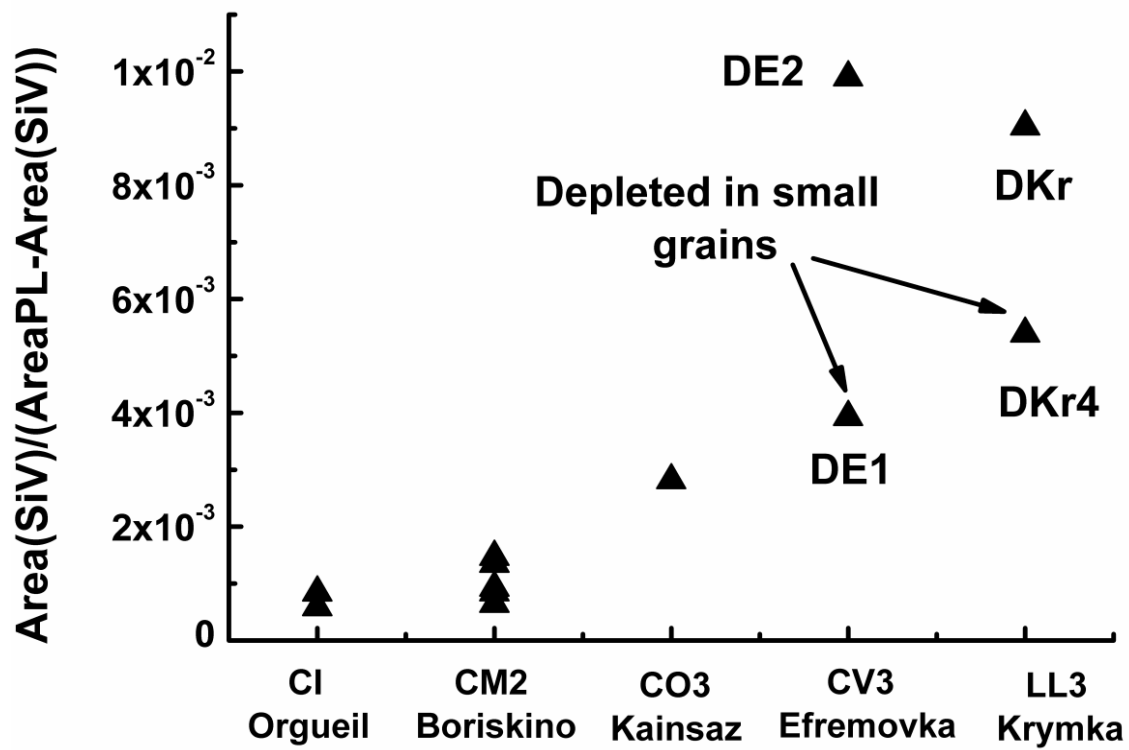

Figure 3. Relative area of the SiV band in dependence of chemical composition of parent meteorite. The total area was calculated between 490 and 800 nm; the SiV band area was calculated after linear background subtraction between 720 and 760 nm. For all samples except Kainsaz points corresponding to size fractions are shown.

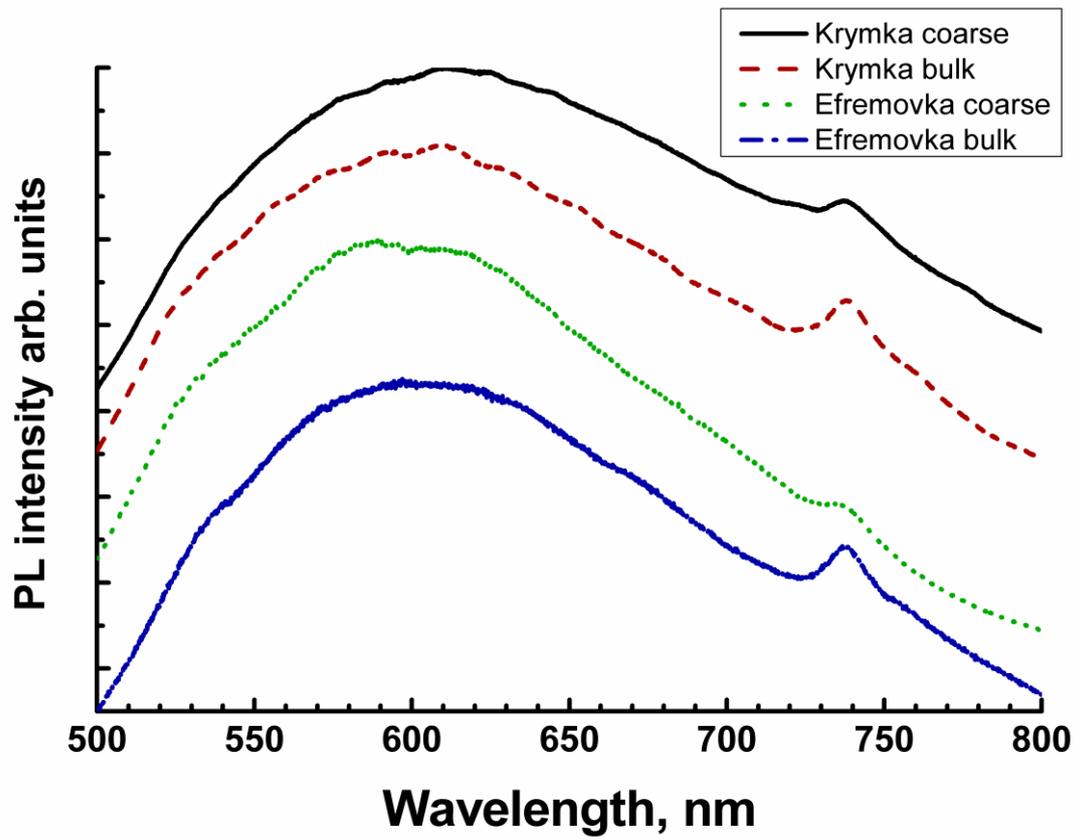

Figure 4. Normalised and vertically displaced spectra of two density fractions of MND from Krymka LL (measured without annealing) and Efremovka CV3 chondrites.